# Generation of abdominal synthetic CTs from 0.35T MR images using generative adversarial networks for MR-only liver radiotherapy

Jie Fu[1], Kamal Singhrao[1], Minsong Cao[1], Victoria Yu[1], Anand P. Santhanam[1], Yingli Yang[1], Minghao Guo[2], Ann C. Raldow[1], Dan Ruan[1], and John H. Lewis[1,3]

1 Department of Radiation Oncology, University of California, Los Angeles, Los Angeles, CA, USA, 90095
2 School of Biomedical Engineering, Shanghai Jiao Tong University, Shanghai, China, 200240
3 Department of Radiation Oncology, Cedars Sinai Medical Center, Los Angeles, CA, USA, 90048
Email: jiefu@mednet.ucla.edu

**Abstract**

Electron density maps must be accurately estimated to achieve valid dose calculation in MR-only radiotherapy. The goal of this study is to assess whether two deep learning models, the conditional generative adversarial network (cGAN) and the cycle-consistent generative adversarial network (cycleGAN), can generate accurate abdominal synthetic CT (sCT) images from 0.35T MR images for MR-only liver radiotherapy.

A retrospective study was performed using CT images and 0.35T MR images of 12 patients with liver (n=8) and non-liver abdominal (n=4) cancer. CT images were deformably registered to the corresponding MR images to generate deformed CT (dCT) images for treatment planning. Both cGAN and cycleGAN were trained using MR and dCT transverse slices. Four-fold cross-validation testing was conducted to generate sCT images for all patients. The HU prediction accuracy was evaluated by voxel-wise similarity metric between each dCT and sCT image for all 12 patients. dCT-based and sCT-based dose distributions were compared using gamma and dose-volume histogram (DVH) metric analysis for 8 liver patients.

$sCT_{cycleGAN}$ achieved the average mean absolute error (MAE) of 94.1 HU, while $sCT_{cGAN}$ achieved 89.8 HU. In both models, the average gamma passing rates within all volumes of interest were higher than 95% using a 2%, 2 mm criterion, and 99% using a 3%, 3 mm criterion. The average differences in the mean dose and DVH metrics were within ±0.6% for the planning target volume and within ±0.15% for evaluated organs in both models.

Results demonstrated that abdominal sCT images generated by both cGAN and cycleGAN achieved accurate dose calculation for 8 liver radiotherapy plans. $sCT_{cGAN}$ images had smaller average MAE and achieved better dose calculation accuracy than $sCT_{cyleGAN}$ images. More abdominal patients will be enrolled in the future to further evaluate two models.

## 1. Introduction

The superior soft tissue contrast of magnetic resonance imaging (MRI), compared to that of computed tomography (CT), allows better tumor and healthy tissue differentiation in certain body areas, such as the brain, pelvis, and abdomen (Khoo and Joon 2006, Schmidt and Payne 2015). MR images are often acquired for tumor and organs at risk (OARs) delineations in treatment planning workflows for pelvic or abdominal cancer radiotherapy (Villeirs *et al* 2005, Lim *et al* 2011, Heerkens *et al* 2017, Mittauer *et al* 2018). Since there is no direct relationship between MR intensity values and electron densities, the standard MR-guided radiotherapy workflow still requires the acquisition of a CT image for dose calculation. However, registration between CT and MR images for transferring target delineations introduces systematic uncertainties that propagate throughout the treatment (Edmund and Nyholm 2017). Acquiring an additional CT image also increases unwanted radiation exposure, clinical workload, and financial cost (Karlsson *et al* 2009). MR-only radiotherapy can avoid these downsides.

A few methods have been proposed to generate synthetic CT (sCT) images from MR images. These methods include atlas-based methods, voxel-based methods, and hybrid methods (Edmund and Nyholm 2017). In atlas-based methods (Sjölund *et al* 2015, Dowling *et al* 2015), the target MR image was first deformably registered to atlas-MR images to acquire deformation vector fields. The acquired vector fields were then reversely applied on the atlas-CT images which were registered to atlas-MR images to generate the sCT image. Atlas-based approaches may not only take a long time to generate the sCT image but also fail if the target patient has substantially different anatomy compared to atlas-patients. Voxel-based methods used machine learning methods that were trained to covert voxel intensities of a single or multiple MR images to CT Hounsfield Units (HUs) (Johansson *et al* 2011, Andreasen *et al* 2016). Hybrid methods combined elements of voxel-based and atlas-based approaches. (Gudur *et al* 2014, Siversson *et al* 2015)

Recently, deep learning (LeCun *et al* 2015), a subset of machine learning, has drawn great research interests for sCT generation mainly due to its fast generation speed and high accuracy. Han (2017) proposed a 2D convolutional neural network (CNN) that achieved accurate brain sCT generation. A study reported that the proposed 2D CNN generated the most accurate pelvic sCT images compared to four atlas-based methods (Arabi *et al* 2018). Fu *et al* (2019) proposed a 3D CNN that generated more accurate pelvic sCT images than Han's 2D CNN. Generative adversarial networks (GANs) were shown to have better performance in image-to-image translation tasks compared to the corresponding CNNs (Isola *et al* 2016, Zhu *et al* 2017). Two popular GANs, the conditional GAN (cGAN) and the cycle-consistent GAN (cycleGAN), were investigated for generating pelvic and brain sCT images, respectively (Wolterink *et al* 2017, Maspero *et al* 2018). Results demonstrated that cGAN and cycleGAN could generate accurate pelvic and brain sCT images, respectively. The pelvic sCT images generated by the cGAN achieved accurate dose calculation for pelvic radiotherapy (Maspero *et al* 2018). A study showed that a modified 3D patch-based cycleGAN generated accurate abdominal sCT images and demonstrated its potential for MR-only liver stereotactic body radiotherapy (SRBT) planning (Liu *et al* 2019). Unlike cGANs, which require co-registered MR-CT image pairs for training, cycleGANs can be trained in an unsupervised manner. This could potentially enlarge the amount of data available for training cycleGANs. So far, no direct comparison of cGAN and cycleGAN for sCT generation has been made.

Although most studies showed that deep learning methods achieved promising performance in generating brain and pelvic sCT images, few studies on the application of deep learning methods to abdominal sCT generation have been published. Larger intra-scan and inter-patient anatomical variations, compared to those in the brain or pelvis, introduce significant challenges in the task of abdominal sCT generation. Interest in low-field MR-guided radiotherapy has grown rapidly in recent years. However, to our knowledge, no deep learning methods have been investigated for generating sCT images from low-field MR images. Compared to high-field MR images, low-field MR images have lower signal-to-noise ratios and more image artifacts caused by lower magnetic field homogeneity. This may result in poor image quality of the sCT images generated by deep learning models.

This study provides the first investigation on applying deep learning methods for generating abdominal sCT images from low-field MR images in support of MR-only liver radiotherapy. We trained cGANs and cycleGANs to generate sCT images from 0.35 T abdominal MR images. sCT HU accuracy was evaluated using voxel-based metrics. sCT-based dose calculation accuracy was also evaluated for liver cancer patients.

## 2. Materials and methods

2.1 Dataset

This study was conducted using 12 abdominal cancer patients (8 liver and 4 non-liver) who underwent MR-guided radiotherapy. Table 1 summarizes the patient characteristics and prescribed doses. All patients had MR and CT images acquired before the treatment. MR images were acquired with a true fast imaging with steady-state precession (TrueFISP) sequence on a 0.35T MRI scanner of the MRIdian system (ViewRay, OH, USA) during 25 s breath hold. MR slice thickness was 3 mm and in-plane resolution was $1.5 \times 1.5$ mm$^2$. Breath-hold CT images were acquired on a 16-slice CT scanner (Sensation Open, Siemens Medical Solutions, Erlangen, Germany) using 120 kVp and 360 mA. CT slice thickness was 1.5 mm and in-plane resolution was $0.98 \times 0.98$ mm$^2$. The target and OARs were delineated by radiation oncologists and medical physicists on the MR images. CT images were deformably aligned to MR images in the MRIdian treatment planning system to create deformed CT (dCT) images for treatment planning.

| Tumor location | Age | Gender | Tumor volume [cc] | Total dose [Gy] | No. of fraction |
|---|---|---|---|---|---|
| Liver | 36 | Female | 23.6 | 45 | 3 |
| | 43 | Male | 44.1 | 60 | 3 |
| | 47 | Female | 1493.4 | 42 | 15 |
| | 54 | Female | 56.3 | 40 | 5 |
| | 54 | Female | 981.4 | 50 | 10 |
| | 55 | Male | 19.3 | 50 | 10 |
| | 58 | Female | 13.5 | 54 | 3 |
| | 71 | Female | 31.5 | 54 | 3 |
| Pancreas | 64 | Male | 25.7 | 40 | 8 |
| Adrenal gland | 71 | Female | 101.6 | 50 | 5 |
| Middle abdomen | 60 | Male | 534.7 | 50.4 | 28 |
| | 70 | Female | 132.6 | 40 | 20 |

Table 1. Patient characteristics and prescribed doses

2.2 Generative adversarial networks

We implemented two generative adversarial networks, cGAN and cycleGAN, for abdominal sCT generation. Figure 1 shows the simplified architecture of the two networks.

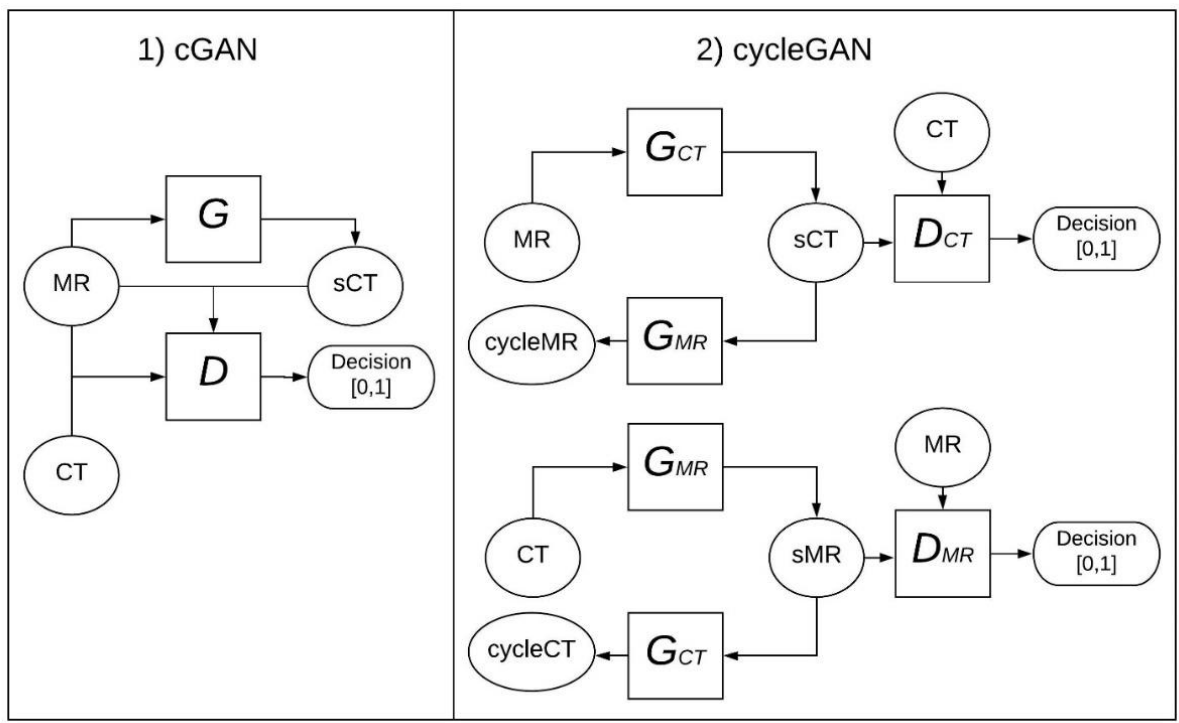

Figure 1. Simplified view of the cGAN and cycleGAN architectures

The cGAN consisted of two convolutional neural networks: a generator (*G*) and a discriminator (*D*). *G* was trained to convert MR slices to sCT slices, while *D* was trained to distinguish the concatenated CT-MR slices from the concatenated sCT-MR slices. The adversarial goal was to generate sCT slices which not only had small L1 distance from CT slices but also could fool *D*. This network requires paired MR and CT slices for training.

The cycleGAN consisted of four CNNs: two generators ($G_{CT}$ and $G_{MR}$) and two discriminators ($D_{CT}$ and $D_{MR}$). $G_{CT}$ ($G_{MR}$) was trained to convert MR (CT) slices to sCT (sMR) slices, and convert generated sMR (sCT) slices back to cycleCT (cycleMR) slices. $D_{CT}$ ($D_{MR}$) was trained to distinguish real CT (MR) slices from sCT (sMR) slices. Unlike the cGAN, the cycleGAN was designed for unsupervised learning, i.e. training with unpaired MR and CT slices in this case. As L1 distance between unpaired CT and sCT slices is not valid, the adversarial goal is to generate cycleCT (cycleMR) slices that had small L1 distance from CT (MR) slices.

Both cGAN and cycleGAN could be trained to convert MR slices to sCT slices. Since our main goal is to test the feasibility of generating abdominal sCT images using GANs, we implemented the same network architectures presented by Isola *et al* (2016) and Zhu *et al* (2017). The networks were modified to process and generate 16-bit single channel images.

2.3 sCT generation

N4 bias field correction (Tustison *et al* 2010) was applied to all MR images to remove inhomogeneity artifacts. Histogram-based intensity normalization (Nyul *et al* 2000) was then performed to minimize the inter-patient MR intensity variation. MR voxel intensities were clipped within the interval [0, 99$^{th}$ percentile], and dCT voxel intensities were clipped within the interval [-1000,1200] HU.

Four-fold cross-validation testing was conducted to generate sCT images for all 12 patients. The patient cohort was randomly divided into four groups. Three groups of 3 patients were used to train the network, the trained network was then applied on the MR images of the patients in the remaining group to generate their sCT images. The cGAN was trained with paired transverse MR and dCT slices, while the cycleGAN was trained with unpaired transverse MR and dCT slices. We adopted the same training protocols presented by Isola *et al* (2016) and Zhu *et al* (2017) for training cGAN and cycleGAN, respectively. Both models were implemented using Tensorflow (Abadi *et al* 2016) packages (V1.3.0, Python 2.7, CUDA 8.0) on Ubuntu 16.04 LTS system, and trained for 200 epochs with a batch size of 1 on a GeForce GTX 1080 Ti GPU (NIVIDIA, California, USA). The L1 loss regularization parameters were set as 100 for training.

2.4 sCT evaluation

dCT and sCT image similarity was evaluated using mean absolute error (MAE) and peak-signal-to-noise-ratio (PSNR) within the MR body contour.

$$MAE = \frac{1}{N}\sum_{i=1}^{N}|I_{sCT}(i) - I_{dCT}(i)| \text{ and}$$

(1)

$$PSNR = 20 log_{10} \frac{4095}{\sqrt{\frac{\sum_{i=1}^{N}(I_{sCT}(i)-I_{dCT}(i))^2}{N}}},$$

(2)

where N is the number of voxels inside the MR body contour, and $I_{sCT}(i)$ and $I_{dCT}(i)_i$ represent the HU values of the i[th] voxel in the sCT and dCT, respectively. In general, lower MAE values and higher PSNR values indicate higher HU prediction accuracy.

Dosimetric evaluation was conducted using clinical plans from 8 liver patients. All plans were optimized on the dCT images according to the clinical guideline using the MRIdian treatment planning system. Dose distributions were calculated using the planning system's Monte Carlo algorithm with magnetic field corrections included. Clinical plans were copied to the corresponding sCT images, and the dose was recalculated using the same calculation protocol. dCT-based and sCT-based dose distributions were compared by gamma analysis (Low *et al* 1998) at 2%, 2mm and 3%, and 3mm within the volumes receiving at least 30%, 60%, and 90% of the prescribed dose. Mean dose and other clinically relevant dose-volume histogram (DVH) metrics were evaluated for the planning target volume (PTV) and OARs. Percentage differences ($\frac{D_{sCT}-D_{dCT}}{D\_prescribed}$) between these metrics calculated with sCT and dCT plans were computed.

## 3. Results

It took about 3 (15) hours to train an individual cross-validation cGAN (cycleGAN). On average, the time required for generating the sCT image of one patient was about 6 s using either model. Figure 2 shows the generated sCT slices along with corresponding MR and dCT slices from 3 representative liver cancer patients. Visual inspection reveals that sCT images generated by the cycleGAN are sharper than those generated by the cGAN. Both models achieved adequate performance in predicting HU values of air pockets, vertebral bodies, and soft tissues but had difficulties in reproducing ribs.

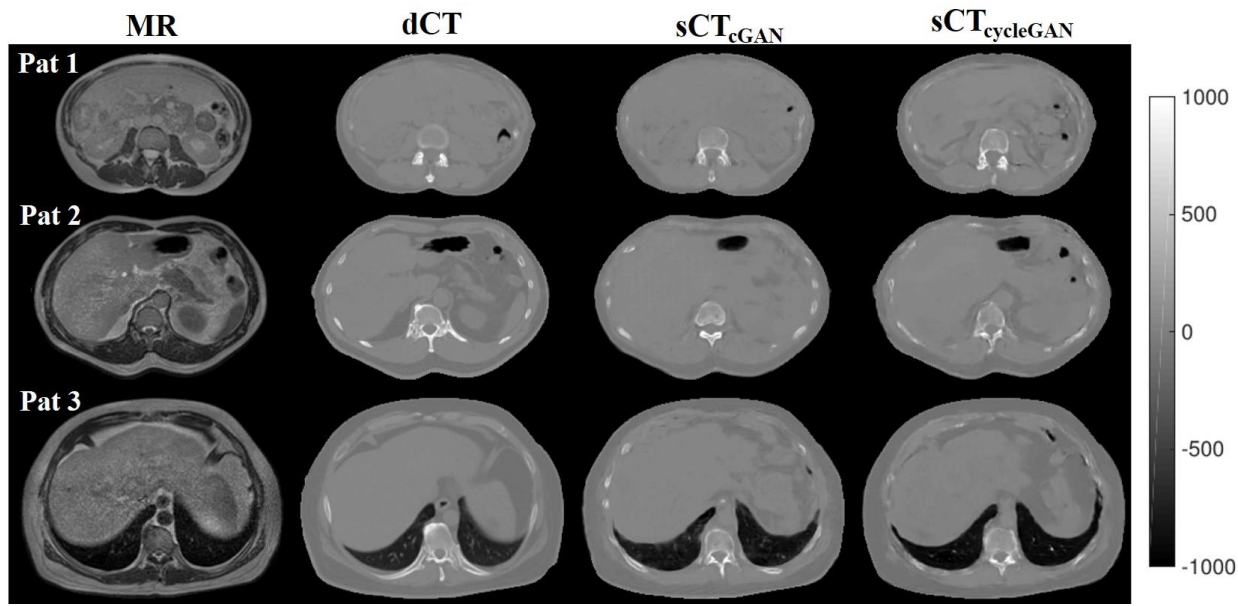

Figure 2. Transverse slices of the MR, dCT, sCT$_{cGAN}$, and sCT$_{cycleGAN}$ images from 3 liver cancer patients. The gray scale bar indicates the HU scale of the CT slices.

For all 12 abdominal cancer patients, MAEs and PSNRs between dCT and sCT images were computed using Eq. (1)-(2). The statistics are summarized in Table 2. On average, cGAN achieved smaller MAE and higher PNSR compared to the cycleGAN. The small patient number resulted in large standard deviations.

|  | MAE [HU] | PSNR [dB] |
|---|---|---|
| cGAN | 89.8±18.7 | 27.4±1.6 |
| cycleGAN | 94.1±30.0 | 27.2±2.2 |

Table 2. Statistics of MAE and PNSR between dCT and sCT images generated by cGAN or cycleGAN. Results were averaged across all 12 patients and shown in (mean ± SD) format.

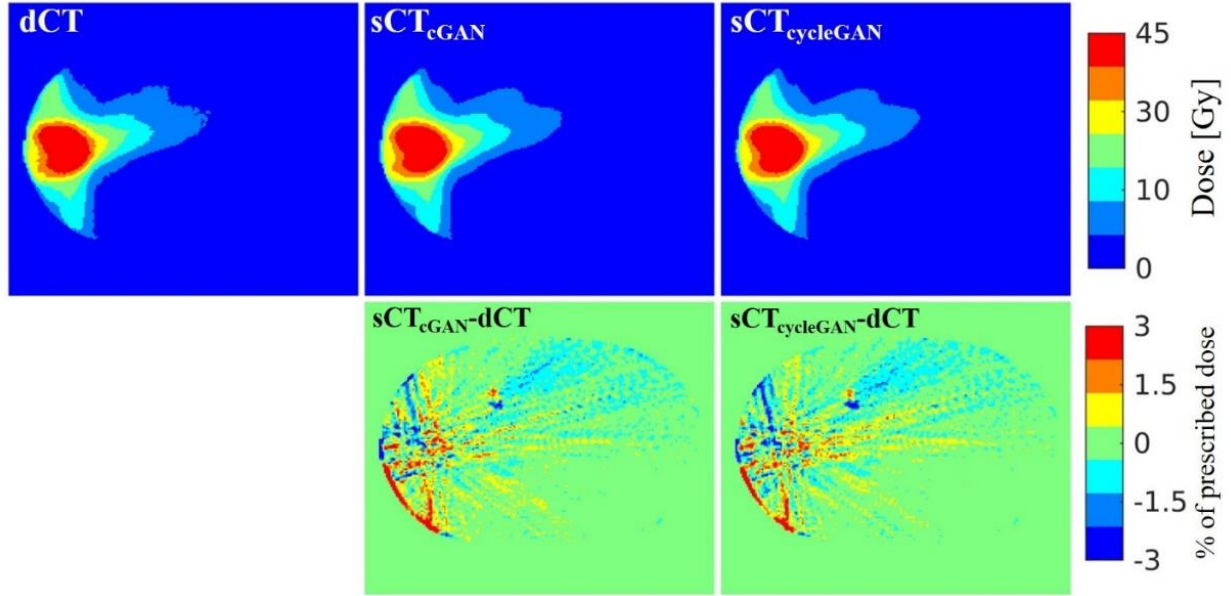

Figure 3. Comparison of dCT-based, sCT$_{cGAN}$-based, and sCT$_{cycleGAN}$-based dose distribution. The first column shows the transverse slices of three dose distributions for one liver cancer patient, and the corresponding dose difference maps between sCT and CT are presented in the second column as the percentage of the prescribed dose (45 Gy).

For 8 liver cancer patients, the clinical plans optimized on dCT images were recalculated with the corresponding sCT$_{cGAN}$ and sCT$_{cycleGAN}$ images, respectively. Figure 3 shows transverse slices of the dCT-based, sCT$_{cGAN}$-based, and sCT$_{cycleGAN}$ dose distributions along with corresponding difference maps for one liver patient. Both sCT images yielded dose distributions that were very similar to those calculated with the dCT image. The dCT-based and sCT-based dose distributions were compared using gamma analysis for the two models. As shown in Table 3, the average gamma passing rates within all evaluated volumes were above 95% using 2%, 2 mm criterion, and 99% using a 3%, 3 mm criterion in both models. The sCT$_{cGAN}$ plan achieved higher average gamma passing rates using a 2%, 2m criterion than the sCT$_{cycleGAN}$ plan.

| Gamma passing rate | Volume of interest | cGAN | cycleGAN |
| --- | --- | --- | --- |
| $\gamma_{3\%,3mm}$ [%] | D ≥ 30% | 99.5±0.8 | 99.5±0.7 |
|  | D ≥ 60% | 99.6±0.7 | 99.6±0.9 |
|  | D ≥ 90% | 99.5±1.1 | 99.3±1.3 |
| $\gamma_{2\%,2mm}$ [%] | D ≥ 30% | 98.7±1.5 | 98.5±1.6 |
|  | D ≥ 60% | 98.4±2.2 | 97.6±2.8 |
|  | D ≥ 90% | 97.4±3.2 | 95.6±5.0 |

Table 3. Statistics of gamma passing rates within the volumes of interest. Results were averaged across 8 liver cancer patients and shown in (mean ± SD) format.

Mean doses and clinically relevant DVH metrics for the PTV and OARs were computed for dCT and sCT plans. Deviations of these metrics between dCT and sCT plans are shown in Table 4. In both models, the average deviations of all metrics were small, within ±0.6% for the PTV and within ±0.15% for all evaluated OARs. Figure 4 presents mean dose differences of the PTV and OARs for all 8 liver cancer patients. The maximum absolute differences of PTV mean doses were 0.4% for the cGAN and 1.0% for the cycleGAN. The cGAN achieved smaller deviation ranges (maximum-minimum) than the cycleGAN for all evaluated regions except the right kidney.

| Regions | Metric | cGAN Deviation sCT vs dCT (% of prescribed dose or volume percentage difference) | cycleGAN Deviation sCT vs dCT (% of prescribed dose or volume percentage difference) |
|---|---|---|---|
| PTV | Mean | -0.17±0.22 | 0.09±0.46 |
|  | $D_{98\%}$ | -0.30±0.31 | -0.06±0.56 |
|  | $D_{95\%}$ | -0.51±0.52 | -0.09±0.94 |
|  | $D_{50\%}$ | -0.27±0.48 | 0.19±0.81 |
|  | $D_{2\%}$ | -0.39±0.63 | 0.17±0.99 |
| Bowel | Mean | -0.05±0.05 | 0.00±0.06 |
|  | $V_{35Gy}$ | -0.05±0.15 | -0.02±0.05 |
| Cord | Mean | 0.04±0.21 | 0.06±0.24 |
|  | Maximum | -0.03±0.29 | 0.01±0.17 |
| Liver | Mean | -0.01±0.10 | 0.03±0.15 |
|  | $D_{1000cc}$ | 0.02±0.08 | 0.06±0.12 |
| Left kidney | Mean | 0.02±0.05 | 0.02±0.05 |
|  | Maximum | -0.08±0.14 | -0.03±0.20 |
| Right kidney | Mean | 0.00±0.08 | 0.01±0.07 |
|  | Maximum | -0.03±0.30 | 0.07±0.34 |
| Stomach | Mean | -0.01±0.03 | -0.05±0.12 |
|  | $V_{35Gy}$ | 0.03±0.09 | 0.15±0.43 |

Table 4. Statistics of metric differences between between dCT and sCT plans. Differences are presented in percentage of prescribed dose (mean, maximum, $D_{2\%}$, $D_{50\%}$, $D_{95\%}$, $D_{98\%}$, $D_{1000cc}$) or volume percentage difference ($V_{35Gy}$). DVH metrics were chosen based on planning constraints for MR-guided stereotactic body radiation therapy requested by physicians. Results were averaged across 8 liver cancer patients and shown in (mean ± SD) format.

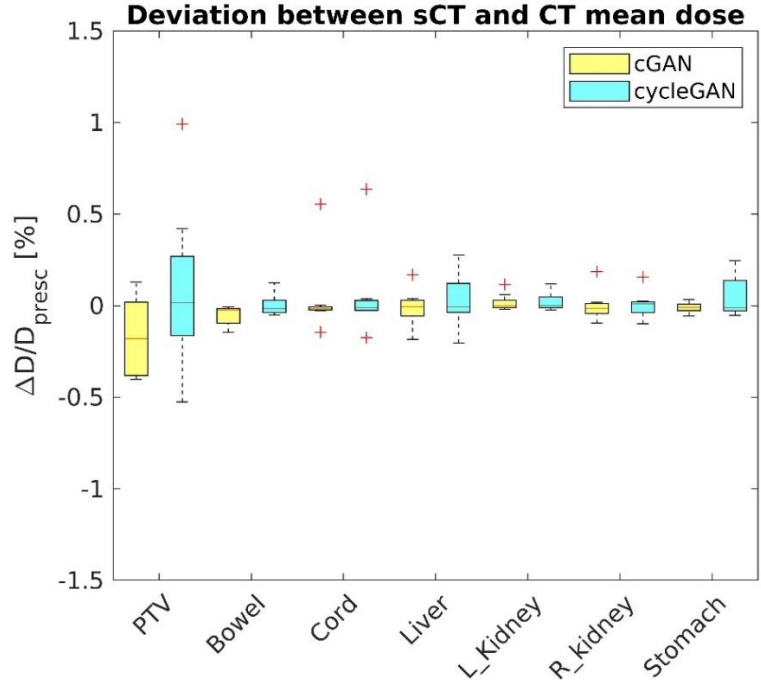

Figure 4. Box and whisker plot of deviations between sCT and CT mean dose within the PTV and OARs. The maximum (top line), 75% (top of box), median (central line), 25% (bottom of box), and minimum (bottom line) are shown. Outliers are drawn as red cross signs. cGAN and cycleGAN results are presented in yellow and cyan, respectively.

## 4. Discussion

In this study, for the first time, deep learning methods have been applied for generating abdominal sCTs images from low-field MR images. We trained two deep learning models, cGAN and cycleGAN, to generate sCT images from 0.35T MR images for 12 abdominal cancer patients. sCT HU accuracy was evaluated using voxel-wise metrics. To evaluate the sCT dose calculation accuracy for liver radiotherapy, we compared the dCT-based and sCT-based dose distributions for 8 liver cancer patients. In both models, the average MAE between dCT and sCT images is of similar magnitude to that previously reported for sCT abdominal generation using high-field MR images (Liu *et al* 2019). The average gamma passing rates were above 99% using a 3%, 3 mm criterion in both models. Small deviations in the mean dose and clinically relevant DVH metrics between sCT- and dCT-based dose distributions were observed in both models as shown in Table 4. These results suggested that abdominal sCT images generated by cGAN or cycleGAN achieved accurate dose calculation for liver radiotherapy planning in our patient-cohort.

Our results also showed that $sCT_{cGAN}$ images had smaller average MAEs and higher average gamma passing rates (using a 2%, 2mm criterion) than $sCT_{cycleGAN}$ images. More patients are required to conduct meaningful statistical tests. Model performance may be limited by the small training dataset. A larger training dataset may lead to more accurate and robust model performance. For example, a deep learning model trained with patients having only small inter-patient anatomical variations may have difficulty in providing accurate HU prediction for patients with atypical anatomies. More liver cancer patients will be enrolled in the future to improve model performance and further compare cGAN and cycleGAN.

The average sCT generation time is less than 10 s using either cGAN or cycleGAN. Generation time can be affected by several factors including image dimension and GPU model. A phase I trial study suggested that stereotactic MR-guided online adaptive abdominal radiotherapy allowed PTV dose escalation and simultaneous OAR sparing compared to non-adaptive treatment (Henke *et al* 2018). The fast sCT generation speed of cGAN and cycleGAN makes it possible to achieve MR-only online adaptive workflow without extensively elongating the time required to adapt.

Deep learning models investigated in this work can also be trained to convert high-field MR images to sCT images. Using high-field MR images may result in better sCT quality since high-field MR images have higher signal-to-noise ratios and fewer image artifacts related to the low magnetic homogeneity than low-field MR images. Future work includes acquiring high-field MR images and investigating the dose calculation accuracy of the sCT images generated from high-field MR images using other commercial treatment planning systems.

**Conclusion**

We implemented cGAN and cycleGAN to generate abdominal sCT images from 0.35T MR images. In this preliminary study, sCT images generated by both models enabled accurate dose calculations for liver radiotherapy planning. The fast generation speed and high dose calculation accuracy make both GANs promising tools for MR-only liver radiotherapy. More abdominal cancer patients will be enrolled in the future to further compare the dose calculation accuracy of the sCT images generated by cGAN and cycleGAN.

**Acknowledgement**
This research was partially funded by Varian Medical Systems, inc.

**Disclosure of Conflicts of Interest**
The authors have no relevant conflicts of interest to disclose.